\documentstyle[12pt]{article}
\begin{document}
\begin{center}
{\Large  Entropic issues in contemporary cosmology}\\
\bigskip
{\bf D.H. Coule}\\
\bigskip
Institute of Cosmology and Gravitation,\\ University of
Portsmouth, Mercantile House, Portsmouth PO1 2EG.\\
\bigskip

\begin{abstract}

Penrose [1] has emphasized how the initial big bang singularity
requires a special low entropy state. We address how  recent brane
cosmological schemes address this problem and whether they offer
any apparent resolution. Pushing the start time back to
$t=-\infty$, or utilizing maximally symmetric AdS spaces, simply
exacerbates or transfers the problem.
 Because the entropy of de Sitter space
is $S\leq 1/\Lambda$, using the present acceleration of the
universe as a low energy $(\Lambda\sim 10^{-120}$) inflationary
stage, as in cyclic ekpyrotic models, produces a gravitational
heat death after one cycle. Only higher energy driven inflation,
together with a suitable, quantum gravity holography style,
restriction on {\em ab initio} degrees of freedom, gives a
suitable low entropy initial state. We question the suggestion
that a high energy inflationary stage could be naturally reentered
by Poincare recurrence within a finite causal region of an
accelerating universe.

 We further give a heuristic  argument that
so-called eternal inflation is not consistent with the 2nd law of
thermodynamics within a causal patch.
\\

PACS numbers: 04.20, 98.80.Hw
\\
Key Words: entropy, inflation, branes, cyclic ekpyrotic cosmology

\end{abstract}
\end{center}
\newpage
{\bf 1. Introduction}

When cosmological models are extrapolated back in time they
generally reach an epoch when  quantum gravity is necessary. This
fortuitously allows us to sweep away many outstanding problems of
the big bang theory and await their resolution when the correct
theory is available. There is also the hope that within such a
theory the ``entrance'' to the universe will become apparent.  Of
course we are too impatient to await an ultimate theory, and
believing cosmology can provide feedback to this search, we set
goals or define problems to be resolved.

 The other side of the coin is that almost anything
  can be allowed during this unknown
quantum gravity region and this is now causing a crisis of
possibilities. For example, the pre-big bang [3] and ekpyrotic
models [4] push the start time back to minus infinity. They
therefore almost to incredible accuracy satisfy the Perfect
Cosmological Principle (see eg.[5]) over huge epochs : the
universe behaves mostly through its lifetime like the steady-state
universe (see [6] for review) which was earlier believed
discounted. Other approaches allow the fundamental constants to
vary [7] and so confuse what puzzles of cosmology are actually
paramount [8]. For example the holography principle [9] suggests
only one degree of freedom per Planck area $\hbar G/c^3$. But note
how this is altered, for example, in models where $c$ initially
tends to infinity the initial number of degrees of freedom becomes
potentially unbounded as the Planck area vanishes.

In this paper we wish to consider how these new, and apparently
radical, cosmological models address a fundamental question. Why
is entropy within the universe able to increase and so allow the
generalized principle of the 2nd law of thermodynamics [10] to
arise? Such arguments can help decide whether the models are at
least plausible, or if the  problems are simply transferred into
understanding the initial state.

  Penrose has suggested that
the Weyl curvature tensor could be related to the gravitational
entropy [1]. Although counter examples to this idea are known
since some anisotropic models evolve to gravitational waves with
zero Weyl curvature [11], it does give some idea of how some
singularities can be more ordered than others. The usual big bang
singularity presumably having small entropy or negligible  Weyl
curvature while  a big crunch could have large entropy or a
dominant Weyl component. The actual definition of gravitational
entropy will not be so important for our discussion since it is
believed known how to define it when cosmological horizons are
present, in for example, de Sitter space.

{\bf 2.0  Standard Inflation}

There is an interesting property of de Sitter space: the higher
its temperature the lower is the entropy. This allows a high
energy cosmological constant $\Lambda$, or equivalent  scalar
potential $V(\phi)$, to still have low entropy. Since [12]
\begin{equation}
S\leq\frac{1}{\Lambda}\simeq \frac{1}{V(\phi)}
\end{equation}
where the temperature is given by
\begin{equation}
T=\Lambda^{1/2} \simeq V(\phi)^{1/2}
\end{equation}
Strictly speaking this is the entropy within the causal  event
horizon. This only agrees with the total entropy of the space at
the minimum size of a closed de Sitter universe. The total entropy
can be infinite in the non-closed case, unless one suitably
compactifies the metric -see eg[13]. This de Sitter case can be
contrasted with the case of gravity being simply attractive. For
example, larger Black holes which maximize the entropy  have lower
temperatures [14].

This example of holography, that bounds the entropy of de Sitter,
prevents a singularity being present or else would allow
violations in  the generalized 2nd law cf.[15] . Therefore, if
such principle, is valid it has a drastic consequence for
inflation: it means that degrees of freedom are being restricted
{\em ab initio} with  an increasing Planck sized $V(\phi)$
potential.  The ``no hair'' property of inflation (see eg. [16])
is implemented automatically. This is helpful since in a classical
analysis it is known that any upper bound in the potential can
lead to ambiguous predictions from inflation: the flatness would
still be arbitrary [17-18] using a classical canonical measure
[19]: essentially because the kinetic energy $\dot{\phi}^2$ can
diverge.  To state this differently: without holography there is
potentially an infinite number of extra degrees of freedom that
could be added within a Planck volume cf. [20]. These cannot be
smoothed by a finite amount of inflation. This resolves a problem
outlined by Penrose [2], that if the quantum gravity region was
``fractal'' inflation would just transfer this fuzziness to large
scales. In contrast if a large cosmological constant already
saturates the holography bound it naturally keeps the space-time
smooth and of low entropy. If one conversely tried to maximize the
entropy then we would obtain the opposite i.e. $\Lambda
\rightarrow 0$ , and indeed people have argued that this could be
a reason why the cosmological constant is so small today [21].
This argument is however deficient since maximizing the entropy
would not allow further increase for the 2nd law to hold.

 Can a large $V(\phi)$ be justified? Somewhat counterintuitive
 the minimum size of closed de Sitter space is $\sim
 \Lambda^{-1/2}$ , so that a smaller size corresponds to larger
 $\Lambda$, or scalar potential $V(\phi)$. In the limit
 $V(\phi)\rightarrow 1$ the minimum size approaches the Planck
 length. Using quantum cosmology one envisions that tunneling to
 this minimum from zero size can occur[22]\footnote{Note that
  while the total entropy of the space will subsequently increase during the
  expansion  the total entropy in the  collapsing de Sitter
   phase  is decreasing which
  would apparently violate the 2nd Law of thermodynamics. This helps
  explains the concern
  of Price that deflation should  also be included [23]}
   However, this is rather an
 extravagant extrapolation of usual quantum mechanical reasoning
 since space-time itself is coming into existence unlike the way
 electrons are treated in for example Alpha decay of nuclei. The
closed de Sitter model itself also  suffers from $\em
{fragility}$[24]: if the curvature
 is removed the minimum size limit is removed. Also if a matter
 component is added the universe becomes instead the Lemaitre
 model (see eg.[5])
 which also starts from a singularity at zero scale factor. One
 can still attempt to quantize such models but the results will
 depend upon arbitrary constants that depend upon the relative matter
 components present. Although in some simple models the quantum measure
 for inflation is still found to be more likely than purely classical
 reasoning would suggest [25].

 If instead of quantum conception we wish to use inflation to
 amplify a ``small bang'' universe other problems are apparent.
 Inflationary conditions necessarily have a singularity in their past [26,27]. To be
 consistent the earlier singularity must also have a very low entropy, so
 this would still need an adequate explanation. If the singularity was
 a
 more generic high entropy one it is difficult to envision how a low entropy domain
 that subsequently inflates could develop.
 The gravitational entropy $S_{W}$ from the Weyl component must be
 strongly constrained if not to dominate the  Sitter entropy $S$.
  In summary, inflation still leaves unanswered either the
 quantum mechanism that produces the initial large smooth field $V(\phi)$
 or for the purely classical case what explains the preceding smooth
 singularity. Although the holography principle can help restrict
 degrees of freedom that require smoothing, the notion of a field
 itself is not consistent with  such a principle. This is at present
 a serious limitation of inflationary theory, particularly when
 considered close to the Planck scale.

{\bf 3.0  Low energy inflation}

If inflation is driven by a small value of $\Lambda$ the maximum
possible entropy increases. For example using the present apparent
value of the cosmological constant as an inflationary phase the
corresponding entropy is $S\sim 10^{120}$. This is the maximum
possible entropy allowed within the horizon and agrees with the
value given by Penrose [1] for a Black hole encompassing the total
mass within the present horizon size. This value will only occur
after a further time $t>H^{-1}\sim 10^{60}t_{pl}$, as particle
production saturates this bound.

In any case inflation does not force local gravitating system
apart. The so-called ``no hair'' theorems of de Sitter space
exclude positive curvature [28]. If one requires that  super
massive Black holes should first evaporate it takes approx $\sim
10^{140}t_{pl}$ [14]. After such time scales the universe will
have around the maximum entropy: being composed of Hawking
radiation with photons at a temperature $\sim 10^{-28}K$ [12].
Note that although the present 3K background radiation will become
redshifted below this value, and contribute negligible entropy,
the quantum effects will dominate. Once the maximum entropy is
reached further evolution or life becomes impossible. Although by
using a weaker power-law expansion $a\sim t^p$ with $p>1$, the
Hubble parameter $H\rightarrow 0$ as $t\rightarrow \infty$. This
would allow the maximum allowed entropy to continue growing and
somewhat alleviate a gravitational heat death [29].

 As mentioned,  ambiguities occur when using a
  canonical measure for the typical classical solutions. If the
inflationary potential is bounded above and the energy density can
be bigger than this value ambiguities in the flatness occur
[17,18]. Reducing the maximum height of this inflationary
potential exacerbates the problem further [18]. This is also
symptomatic of arbitrary values of anisotropy or inhomogeneity
that would result if they were introduced into the model [18]. The
upshot of this work is that any finite amount of inflation does
not itself restrain one to a isotropic and flat FRW universe.
Other principles have to be introduced to prevent such unwanted
values, in for example, the curvature. Making the model cyclic
does not help in this regard since the question simply becomes why
this cyclic model and not that one with different properties of
curvature, anisotropy? Also bear in mind that inflation does not
destroy curvature but only dilutes it during the expansionary
phase. If the universe subsequently collapses it will simply
reappear again, with the anisotropy and finally rotation
dominating as the scale factor tends to zero.

 {\bf 4.0  Eternal inflation}

 Once inflation occurs it is suggested that quantum fluctuations
 can always keep a region of the universe inflationary [30]. If one
 tries to extend this argument backwards in time
 to allow an always expanding inflationary universe it apparently fails [31]. This
 is not surprising since only the closed de Sitter model is
 geodesically complete: even violating the weak energy condition
 as in the steady state model does not remove this geodesic
 incompleteness [32].\footnote { There is an attempt [33], which
  closely resembles the Hoyle-Narlikar model [34]
  to patch together two flat de Sitter
 universes to overcome this restriction. But time runs in opposite direction away from
 an infinite null boundary of low entropy connecting the two universes. }
   There is a simpler way to see this concern. The
 greater the amount of inflation the smaller is the initial region that
 expands to form our universe. If this initially falls within the Planck
 region there is no reason to believe that space-time is
 continuous on this scale but rather notions in quantum gravity
 suggest it is discrete. If space has a cut-off
 at the Planck scale then indeed only a finite amount of inflation
 can have happened. Indeed one can argue that space must be fairly
 smooth even at sub-Planckian scales for inflation to  be usable. There
 has been some works that modify the dispersion relations at high
 energy but this is only a small  departure to the assumption
 of smoothness [35].

 We can also tentatively question the eternal inflationary mechanism to the
 future. Firstly, once quantum gravity is present the scale factor becomes
 dimensional together with the velocity $\dot{a}$. As inflation
 continues this velocity increases exponentially  $\dot{a}\sim \exp(Ht) \rightarrow
 \infty $. This could induce quantum effects as the space expands increasingly
 rapidly.

  It further assumes that  quantum fluctuations are present with
 wavelength $\sim H^{-1}$. These cause the potential to grow
  $V(\phi)\rightarrow V(\phi)+\delta V(\phi)$ in certain domains.
   But from expression (1) this means
   the entropy has decreased in violation of
   the 2nd law of thermodynamics. Should
   this ever be allowed even if many domains are coming into existence?
   In general the typical wavelength of Hawking radiation is
 $\sim A^{1/2}$ where $A$ is the area of the
  event horizon. As $A$
 becomes increasingly large is becomes difficult to justify quantum coherence
 being maintained so that a superposition of modes
 can be justified. This
 becomes particularly suspect with low energy inflation where the
 fluctuations being produced are of order the present size of the
 universe $\sim 10^{60} l_p$. One would expect that they should
 decohere by interacting with the environment and that the field
 would behave classically, with no violation of the weak energy condition,
  on large scales.  The universe would
 always roll towards the minimum of the potential and inflation
 would have a finite future duration. Although this seems clearer in the
 low-energy inflationary case it is also possible that this also
 occurs for high energy inflation where the event horizon is now
 only of typical area $\sim 10^{6} l_{pl}^2$. If quantum gravity ``measures'' the
 system cf.[1] the necessary quantum coherence for the eternal mechanism will be
 lost. There is also a contradiction with the holography principle
 as $A\rightarrow l_{pl}^2$. Degrees of freedom are being restricted and so ``no
 hair'' is being allowed. But Hawking radiation, which is simply quantum mechanics,
  should excite all possible particles that are present in the underlying theory.  All
 matter modes' fluctuations should be generated and these could
 provide an environment for destroying coherence. However, too many species make the
 theory unstable to collapse by black hole generation [36]. A related point
 was made in ref.[37] that expected ``zero-point'' fluctuations
  in radiation affected the
 implementation of the Hartle-Hawking [38] scheme in quantum cosmology.

 This argument against eternal inflation can be contrasted with a
 recent argument of Turok[69], who also suggests that the typical
 evolution does not display the eternal mechanism. His argument
 does not seem complete since he first ignores the quantum effects
 on the classical evolution, and then points out the quantum
 component  is anyway subdominant. The argument presented above would
 first decohere the quantum effects into an averaged classical
 evolution, which obeying the 2nd law of thermodynamics within
 any causal domain, will always cause rolling down the potential.

{\bf 5.0  Universe proliferation }

 There is another scheme for providing an infinite number of
 inflationary universes that I find rather suspect. This is the
 idea that when Black holes are created in de Sitter space and after
 evaporating they leave behind disjoint universes [39].

 This is because the Schwarzschild-de Sitter metric has an
 infinite number of repetitions in its Penrose diagram [32].
 But if a black hole forms by gravitational
 collapse in our present universe we don't claim that the left
 hand universe in the Penrose diagram of Schwarzschild is suddenly
 created. We rather claim that the physical Cauchy development of a collapsing
 star excludes this region. Likewise in our present universe that
 appears to be accelerating the formation of black holes does not
 imply that the  other universes in maximally extended Schwarzschild-de Sitter
 suddenly come into existence and suddenly  become relevant for us physically.
 Indeed by collapsing some matter we apparently would violate
 causality by suddenly having created an infinity of universes
 beyond our horizon. The whole present universe would have to
 change throughout in response to universes now appearing at the
 edges. During this time all
 classical mechanics, including general relativity, ceases to be valid and  is
 simply  being violated throughout the universe; an extreme form of
 horizon problem.
 Now, does quantum
  mechanics make this  argument at all reasonable? The reasoning is that metric A
  can become B by simply calculating the relevant actions of the
  two metrics and this gives some probability for going from one
  to another. But this only makes sense if classical behaviour of
  the universe can be entirely suspended. This might be possible
  at the Planck scale but for a closed de Sitter universe the
  total volume of the universe is rapidly becoming larger than the
  Planck volume for times $t>t_{pl}$. Studies in loop quantum geometry
  suggest that space rapidly becomes classical at scales above
  Planck size, see ref. [40] for review.
   This would prevent the sort of proliferation of
   space-time processes outlined
 which although having a small action are not confined to Planck
 length scales.

 {\bf 6.0  Semi-eternal cosmology}

We have in mind cosmologies  that start at time $t=-\infty$ before
simulating a big bang at time $t=0$. Since the singularity at time
$t=0$ has to be smooth (low Weyl curvature) we have to have an
even lower entropy at the initial start $t=-\infty$. One would
expect the entropy to grow significantly during a semi-infinite
interval of time. For example in the pre-big bang model any slight
initial classical perturbations will grow during the collapsing
phase. Likewise in colliding brane schemes like the original
ekpyrotic one [4], the initial branes have to have small entropy
that cannot dominate over the entropy produced by the collisions.

Another example is  an eternal brane produced by achieving a
bounce that prevents a singularity forming. For a
Reissner-Nordstrom AdS bulk the Friedmann equation becomes
modified, such that [41,42]
 \begin{equation}
 H^2+\frac{k}{a^2}=\frac{8\pi G}{3} \left (\rho +\frac{\rho^2}{2\lambda}\right )
 +\frac{M}{a^4}-\frac{Q^2}{a^6}
 \end{equation}

 where $M$ and $Q$ represent the mass and charge of the bulk
 space. In the Friedmann equation the $M$ term behaves like
 radiation while the charge $Q$ violates the weak-energy
condition. For a perfect fluid equation of state
$p=(\gamma-1)\rho$, a bounce can occur for matter softer  than
that of dust, i.e. $\gamma<1$.  Again the  problem still to be
overcome is that classical perturbations will tend to grow rapidly
during the collapsing phases. This has already been investigated
in ref. [43] where the scalar field perturbation are prone to
diverge as the singularity is approached. Even if the bounce can
proceed the initial state's order is difficult to justify and
would require other principles for its explanation.

There are some related attempts to use quantum cosmology to give
an initially large universe that could subsequently collapse
cf.[44]. For reasons already discussed in the previous section
this is difficult to envision. There was an earlier argument that
tried to give a reason for the present value of $\Lambda$ but
implicitly assumed the universe had just suddenly ``quantum
created'' itself into its present size [45]. This could be
discounted by arguing that any possible  Euclidean space-time
structure should be constrained to small scales only [46].

{\bf 7.0 Cyclic ekpyrotic cosmology}

This model [47] tries to employ a low energy driven inflationary
stage as a means of explaining the smooth state required before a
collapse to a bounce can proceed. Although this model is driven by
the behaviour of branes in higher dimensions we will only consider
the usual approximation of  describing it by  a scalar field model
in 4 dimensions. The model uses a scalar potential with a negative
region, but somewhat arbitrarily the weak energy condition is not
allowed to be violated, and enforced by keeping only large and
positive kinetic energy. Apart from the problems inherent in low
energy inflation it also suffers from problems found earlier with
the pre-big bang model [48]. Because the kinetic energy dominates
as the scale factor goes to zero the Planck problem of the usual
big bang model occurs [49]. This is true for any model that
doesn't violate the strong-energy condition and means the scale
factor must be extremely large at the Planck time. The scale
factor gains units of length whenever quantum gravity is
introduced. One can see this by extrapolating back the presently
observable universe adiabatically to the Planck time. Another way
to see this is to consider the holography principle or its weaker
requirement that only one degree of freedom per Planck volume is
allowed. If we are to explain the required entropy today in
radiation $\sim 10^{90}$ we require $a> 10^{30} l_{pl}$ back at
the Planck time. The bounce must occur long before the quantum
scale when string theory might be expected to allow further
unknown phenomena. If we insist on allowing the universe to
approach the actual Planck size the values of the various
quantities e.g. energy density, will rapidly go beyond values that
we have any confidence in describing: the Planck problem [49].

The question of quantum fluctuations has produced much debate and
argument [50]. Fluctuation are produced in expanding models that
have an event horizon. But for a collapsing universe the time
reversal of  a particle horizon is an event horizon [5]: so
perturbations are produced in contracting models with scale factor
$a\sim (-t)^p$ with $0<p<1$. A serious concern is readily
apparent. The usual horizon problem (presence of a particle
horizon) only concerns times below Planck times where the
expansion becomes faster than light. Simply altering the behaviour
of the scale factor within the Planck time of the singularity can
remove entirely the presence of the particle horizon [51].
Likewise in the collapsing universe the event horizon required for
quantum fluctuations depends on the behaviour of the scale factor
at below Planck times. If the scale factor changes its behaviour
before this time strictly speaking an event horizon is never
produced. One should instead be able to define a vacuum state
until a Planck time before the impending singularity that does not
produce particles. We will ignore further this concern and assume
such perturbations are actually produced.

It has  been known for some time that a kinetic energy or stiff
equation of state $p=1/3$ produces a blue spectrum $n=3$ of
fluctuations [51]. If one wishes to produce a scale invariant
$n=0$ spectrum during a collapsing phase one requires a dust
equation of state or $p=2/3$ [52]. This is in contrast to the
expansionary case where only a cosmological constant gives such a
spectrum. Although the kinetic energy will undoubtedly dominate as
the bounce approaches there is a period when the negative
cosmological constant also contributes to the equation of state
and gives  slow contraction $p\rightarrow 0$.\footnote{One might
worry that the apparent speed of sound is unphysical for
$0<p<1/3$, since it corresponds to a super stiff equation of state
$\gamma>2$. If the kinetic energy was constrained
$|V(\phi)|>\dot{\phi}^2$ pole-law collapse with an effective
$\gamma<0$ could result [53]. } In this limit it still produces a
blue spectrum with $n=2$ as does a collapsing radiation $p=1/2$
case [54]. But because of the enormous time scales involved
($10^{140}t_{pl}$ for super massive black holes to evaporate)
without fine tuning, only the fluctuations during the previous
kinetic dominated phase should be relevant for our present early
universe. It is presently no further than a fraction $10^{-80}$ of
its total lifetime between bounces.

Returning to the entropy question. It is suggested that inflation
dilutes the entropy of matter before the collapse occurs. But this
ignores the quantum particle creation that will saturate the  de
Sitter entropy bound. In the language of ref.[55] a gravitational
heat death occurs. It is true that classically  matter is swept
across the horizon but the event horizon radiates an entropy that
represents possible information lost beyond the event horizon. If
the entropy could go to zero as required to reset the universe,it
would mean that entropy can be destroyed by de Sitter space. This
is like a classical black hole with $S=0$ that likewise appears to
have destroyed entropy. But when quantum mechanics is introduced
the entropy of Black holes and of cosmological event horizons is
given by the corresponding $\sim$  (Horizon Area)$/4l_{pl}^2$, so
that the generalized 2nd law still holds .

Once this entropy is produced it is difficult to envision how it
might be  removed, or to allow an entropy gap between
gravitational entropy and matter entropy to develop [55].  The
photons will become increasingly blue shifted as the singularity
is approached. The total entropy within causal contact will be at
least $\sim 10^{120}$ . Before the next universe cycle occurs this
needs to be somehow dissipated or else the photons never be
allowed to gain a high temperature. The authors of ref.[47] seem
to suggest that these photons, become, in the words of ref.[56],
frozen and so no longer contribute dynamically to the entropy.
Having to dissipate such a large entropy, a factor $\sim 10^{30}$
times larger than the present entropy of the observable universe,
would seem at best to be extravagant, if not contrary to the
generalized 2nd law of thermodynamics.

It might be argued that the actual brane scale factors are always
expanding [47] and the problem is only due to the effective
theory. Although this might help alleviate the bounce the entropy
is due to the cosmological constant on the branes themselves. By
the way if the scale factors always grow more than they collapse
all physical scales come from increasingly smaller scales as more
cycles of the model occur. Eventually all scales would come from
below the 11 dimensional Planck scale or other quantum gravity
scale where the theory was no longer valid. The model would still
have a beginning that requires further explanation.

 Unless there
was some sort of fractal spacetime structure that could constantly
be magnified the model breaks down.
 But this
contradicts our present notions of having a fundamental Planck
scale -see eg.[40]. If the collapse time before the next ``big
crunch'' was shortened before the quantum particle creation
saturated the bound, then we still have problems with large black
holes going into the bounce.

Also if the collapsing epochs are only a small disturbance from
mostly expanding behaviour ( as claimed when working with the
brane scale factors )  a geodesic incompleteness theorem could
also be readily obtained cf. [31]. In ref.[47] it is claimed that
such a theorem is not too serious because all particles are
created afresh at each new bounce. But if there is always Hawking
radiation present, as we argue, this reasoning is invalid and
particles cannot be entirely diluted away prior to each bounce.
This continuance of particles is necessary for  a related geodesic
incompleteness argument to be made  and would  mean that somewhere
a beginning to the model's evolution  would still be required.

{\bf 8.0 Cyclic universe by Poincare recurrence?}

Are there other ways in which a cyclic universe might be possible
which do not suffer a continual increase in entropy? In this
section we just wish to restate how difficult such a scenario
appears with our  current knowledge of physics.

For example, it has been suggested [57] that, in the apparently de
Sitter phase the universe is entering, the universe can naturally
reenter a low entropic state, or  high energy inflationary phase,
given sufficient time to provide a suitable fluctuation. This is
because the number of states within the cosmological horizon is
finite $\sim 10^{120}$ and Poincare recurrence should enable the
space to return to any previous state. This is a old idea in
cosmology- see ref.[11] for a discussion, but is generally
discounted because cosmological models are all actually unbounded
[58] .
 We know from studies in ``universe creation in the lab'' that the
 created universe expands not into the original universe but into
 an
entirely new space [59]. A singularity is left behind in the
original universe and the presence of singularities is one way of
evading the possibility of Poincare recurrence [58]. One could
avoid the creation of a singularity if the new inflationary phase
extended into an antitrapped region, so being larger that the
horizon size of the low energy background [26,27]. This is a vast
size and for similar reasons as were discussed in section 5) it is
difficult to believe quantum tunneling could occur over such
scales vastly larger than the Planck scale . One such model  is
described in ref.[60] where quantum bubbles of false vacuum are
created from a low energy inflationary phase. Such bubbles are
dependent on the background spacetime in order not to collapse,
they do not simply supplant the original spacetime as a  finite
cosmological model allowing eternal return would require. Again it
is an example of an unbounded system which escapes possible
Poincare recurrence. Note that these actual examples of possible
inflationary universe generation  are apparently  not consistent
with horizon complementarity or other claimed  sacrosanct
principles which led the authors of ref.[57] to the opposite
conclusion. This is not too surprising since as emphasised in
ref.[61] our present notions of entropy/horizons are not well
established for highly dynamic spacetimes as in the example above
with new universes being produced.

 {\bf 9.0 Conclusions}

Explaining the low initial entropy of the universe still seems far
from being resolved. In inflationary universes only at the minimum
of the closed de sitter $a\sim \cosh t$, is the total
gravitational entropy of the space minimized. The entropy tends to
infinity as $t\rightarrow \pm \infty$. So only by expunging the
collapsing region ( or starting at $t=0$) can we be consistent
with the 2nd law that entropy should only increase. But doing this
means the model needs a
 further constraint  to cause it to start at the minimum. Within
 this explanation lies the answer to why the entropy is initially low.
 Quantum cosmology tries to address this question but it is a vast
 extrapolated from the usual domain of quantum mechanics. Where
 space-time itself is already in existence. The Hartle-Hawking
 proposal [38] also seems to favour maximizing the entropy $\Lambda\rightarrow
 0$, to also produce a low energy inflation -see however [62] for
 alleviating this prediction.

Inflation is believed to smooth space, but because of a holography
or Planck cut-off principle, the number of degrees of freedom
initially present is severely restricted. This helps make
inflation more durable, unlike a purely classical case that leads
to indefinite values in various quantities. But if inflation is
premised on these degrees of freedom being absent it does not
explain the whole story.

 If inflation is driven by a small cosmological constant then even
 within the event horizon  itself, the entropy, taking its de Sitter value,
  becomes too large. Although classically one expects the entropy to be
 approaching zero,  Hawking radiation from the event horizon
  fills the space with  particles. Once
 gravitational and matter entropies are in equilibrium a heat
 death results. Even if the cosmological constant decays away the photons will
 blue shift during collapse producing vastly
  more entropy at any time in the next cycle than at any corresponding  time
  in our present universe. A similar entropy problem to that in the original Tolman
  cyclic model that also  limits the number of allowed cycles -see eg.[11].
  Claiming that this entropy is no longer ``dynamic'' for the next cycle goes against
  the belief  in black hole physics that information should always be recoverable.
  Neither, is there an expansion phase immediately after  inflation that
  would allow an entropy gap, between gravity and radiation,  to again develop cf. [55].

    This is therefore a serious problem with schemes like the cyclic
  ekpyrotic one, that
 utilize low energy inflation in the hope of producing low entropy
 conditions before a subsequent bounce can occur. Some recent
work [63] also suggests that a high energy inflationary stage
should be added to the scheme.

 Other brane schemes that push the start back to time $t=-\infty$
 only give more time for the entropy to grow and so require even
 more orderly initial states. Sometimes they use a bulk space
 which satisfies the Perfect Cosmological Principle so no
 evolution is allowed. It then becomes difficult to explain why
 dynamic
 branes in low entropy states are then produced [42].  Or else the
 problem simply transfers to understanding the low entropic, or
 orderly, bulk space. Some schemes try to use a prior inflationary
 phase within the bulk before bubbles of Anti-de sitter (AdS)
 space are later produced [64,65]. Depending on various arbitrary
 scales entering the scalar potential
  such an inflationary phase is not unstable to bubble
 formation even if a region of negative potential is present [65-67]. Provided the AdS
  ``well''
 is not too deep, in comparison to the
 false vacuum value of the potential, bubbles cannot form without
 energy conservation violation.  If
 such bubbles do actually form they generally collapse and do not expand to
 fill all the previously inflationary region [65]. Most of this
 difficulty   stems  from now having  a  dynamical bulk so that the  full
 maximal symmetry of DeSitter or AdS is no longer present. Instead the symmetry
is at
 most that of Roberston-Walker type : homogeneity and isotropy,
 where a negative cosmological constant simply produces a rapid collapse to a `big
 crunch'
 singularity [42].

The original remarks of Penrose [1,2] that the present
time-symmetric laws appear insufficient to explain the initial
state of the universe still seems  pertinent. Using string
cosmology to give some explanation for the low entropy state seems
a daunting task, but until then our cosmological models are
missing some vital component.
\newpage
 {\bf Acknowledgement}\\ I should like to thank David Wands and Alexei Nesteruk
  for helpful advice.
\newpage

{\bf References}\\
\begin{enumerate}
\item R. Penrose, ``The Emperor's new mind'' (Oxford University
Press, Oxford) 1989
\item R. Penrose, ``Difficulties with inflationary cosmology''
14th Texas symposium, ed. E.J. Fenyves (New York Academy of
Science, New York) 1989.
\item M. Gasperini and G. Veneziano, Astro. Phys. 1 (1993) p.317.
\item J. Khoury, B.A. Ovrut, P.J. Steinhardt and N. Turok, Phys. Rev.
D 64 (2001) p.123522.
\item W. Rindler, ``Essential Relativity 2nd edn.'' (
Springer-Verlag: New York) 1977.
\item F. Hoyle, G. Burbidge and J.V. Narlikar, ``A different
approach to cosmology'' ( Cambridge University Press: Cambridge)
1999.
\item J.W. Moffat, Int. J. Mod. Phys. A 2 (1993) p.351.\\
A. Albrecht and J. Magueijo, Phys. Rev D 59 (1999) p.043516.\\
J.D. Barrow and J. Magueijo, Class. Quant. Grav. 16 (1999) p.1435.
\item D.H. Coule, Mod. Phys. Lett. A 14 (1999) p.2437.
\item G. 't Hooft, preprint gr-qc/9310026.\\
L. Susskind, J. Math. Phys. 36 (1995) p.6377.
\item J.D. Bekenstein, Nuovo Cim. Lett. 4 (1972) p.737.\\
Phys. Rev. D 7 (1973) p.2333.
\item J.D. Barrow and F.J. Tipler, ``The anthropic cosmological
principle'' (Oxford University Press: Oxford) 1986.
\item G.W. Gibbons and S.W. Hawking, Phys. Rev. D 15 (1977) p.
2738.
\item J.P. Luminet, Phys. Rep. 254 (1995) p.135.
\item S.W. Hawking, Com. Math. Phys. 43 (1975) p.199.
\item J.D. Bekenstein, Acta. Phys. Polon. B 32 (2001) P.3555.
\item D.S. Goldwirth and T. Piran, Phys. Rep. 214 (1992) p.223.
\item D.N. Page, Phys. Rev. D 36 (1987) p.1607.
\item D.H. Coule, Class. Quant. Grav. 12 (1995) p.455.
\item G.W. Gibbons, S.W. Hawking and J.M. Stewart, Nucl. Phys. B
281 (1987) p.736.\\ S.W. Hawking and D.N. Page, Nucl. Phys. B 298
(1988) p.736.
\item F. Englert, preprint hep-th/9911185
\item P. Horava and D. Minic, Phys. Rev. Lett. 85 (2000) p.1610.
\item A. Vilenkin, Phys. Rev. D 30 (1984) p.509.\\
A.D. Linde, Sov. Phys. JEPT 60 (1984) p.211.\\ V.A. Rubakov, Phys.
Lett. B 148 (1984) p.280.\\ Y.B. Zeldovich and A.A. Starobinsky,
Sov. Astron. Lett. 10 (1984) p.135.
\item H. Price, ``Time's arrow and Archimedes' point'' (Oxford
University Press: Oxford) 1996.

\item R.K. Tavakol and G.F.R. Ellis, Phys. Lett. A 130 (1988)
p.217.\\
 A.A. Coley and R.K. Tavakol, Gen. Rel. Grav. 24 (1992)
p.835.
\item D.H. Coule and J. Martin, Phys. Rev. D 61 (2000) p.063501.
\item T. Vachaspati and M. Trodden, Phys. Rev. D 61 (2000) p.
023502.
\item D.H. Coule, Phys. Rev. D 62 (2000) p.124010.
\item R.M. Wald, Phys. Rev. D 28 (1983) p.2118.\\
I. Moss and V. Sahni, Phys. Lett. B 178 (1986) p.159.
\item E. Witten, preprint hep-th/0106109.
\item A.D. Linde, Phys. Lett. B 175 (1986) p.395.
\item A. Borde, A.H.  Guth and A. Vilenkin, preprint gr-qc/0110012
\item S.W. Hawking and G.F.R. Ellis, ``The large scale
structure of space-time'', (Cambridge University press: Cambridge)
1973.
\item A. Aguirre and S. Gratton, Phys. Rev. D 65 (2002) p.083507.
\item F. Hoyle and J.V. Narlikar, Proc. Roy. Soc. A 277 (1964)
p.1.
\item R.H. Brandenberger and J. Martin, Mod. Phys. Lett. A 16 (2001) p.999.\\
 A.A. Starobinsky, JEPT Lett. 73 (2001) p.371.
\item R. Brustein, D. Eichler and S. Foffa, Phys. Rev. D 65 (2002)
p.105013.
\item J.R. Gott and Li-Xin Li, Phys. Rev. D 58 (1998) p.023501.
\item J.B. Hartle and S.W. Hawking, Phys. Rev D 28 (1983) p.2960.
\item R. Bousso, Phys. Rev. D 60 (1999) p.063503.\\
Phys. Rev. D 58 (1998) p.083503.
\item A. Ashtekar, preprint math-ph/0202008 {\em and references
therein}.
\item C. Barcelo and M. Visser, Phys. Lett. B 482 (2000) p.183.\\
 preprint hep-th/0004056.
\item D.H. Coule, Class. Quant. Grav. 18 (2001) p.4265.
\item A.B. Batista, J.C. Fabris and S.V.B. Goncalves, Class.
Quant. Grav. 18 (2001) p.1389.
\item D. Green and W.G. Unruh, preprint gr-qc/0206068.
\item A. Strominger, Nucl. Phys. B 319 (1989) p.722.
\item D.H. Coule, Mod. Phys. Lett. A 10 (1995) p.1989.
\item P.J. Steinhardt and N. Turok, Phys. Rev. D 65 (2002) p.126003.
\item D.H. Coule, Class. Quant. Grav. 15 (1998) p.2803.
\item Y.B. Zeldovich, ``My universe: selected reviews'' (Harwood
Academic Press) 1992 p.95
\item D.H. Lyth, Phys. Lett. B 524 (2002) p.1.\\
R. Brandenberger and F. Finelli, JHEP 0111 (2001) p.056.\\ J.
Hwang, Phys. Rev. D 65 (2002) p.063514.\\ J. Martin, P. Peter, N.
Pinto Neto and D.J. Schwarz, Phys. Rev. D 65 (2002) p.123513.
\item T. Padmanabhan, ``structure formation in the
universe'',(Cambridge University Press: Cambridge) 1993 p.359.
\item Y.B. Zeldovich and I.D. Novikov, ``The structure and
evolution of the universe: relativistic astrophysics vol.2''
(Chicago University Press: Chicago)1983 p.666.
\item D.H. Coule, Phys. Lett. B 450 (1999) p.48.
\item D. Wands, Phys. Rev. D 60 (1999) p.023507.
\item J.D. Barrow, New. Astron. 4 (1999) p.333.\\see also P.C.W.
Davies, in  ``Physical origins of time asymmetry'', eds. J.J.
Halliwell, J. Perez-Mercader and W.H. Zurek, (Cambridge University
Press; Cambridge ) 1994.
\item R. Brustein, Phys. Rev. Lett. 84 (2000) p.2072.
\item L. Dyson, M. Kleban and L. Susskind, preprint
hep-th/0208013.\\ see also D. Bak, preprint hep-th/0208046.
\item F.J. Tipler, Nature 280 (1979) p.203.
\item S. Blau, E. Guendelman and A.H. Guth, Phys. Rev D 35 (1987)
p.1747.\\ E. Fahri and A.H. Guth, Phys. Lett. B 183 (1987)
p.149.\\ E. Fahri, A.H. Guth and J. Guven, Nucl. Phys. B 339
(1990) p.417.
\item J. Garriga and A. Vilenkin, Phys. Rev. D 57 (1998) p.2230.
\item A. Corichi and D. Sudarsky, Mod. Phys. Lett. A 17 (2002)
p.1431.
\item D.N. Page, Phys. Rev. D 56 (1997) p.2065.
\item G. Felder, A. Frolov, L. Kofman and A. Linde, Phys. Rev. D
66 (2002) p.023507.
\item M. Bucher,preprint hep-th/0107148.
\item U. Gen, A. Ishibashi and T. Tanaka, Phys. Rev. D 66 (2002)
p.023519.
\item S. Coleman and F. De Luccia, Phys. Rev. D 21 (1980) p.3305.
\item S. Weinberg, Phys. Rev. Lett. 48 (1982) p.1776.
\item V.A. Berezin, V.A. Kuzmin and I.I. Tkachev, Phys. Rev. D
36 (1987) p.2919.
\item N. Turok, Class. Quant. Grav. 19 (2002) p.3449.
\end{enumerate}
\end{document}